\documentstyle[12pt]{article}

\newtheorem{prop}{Proposition}
\newtheorem{cor}{Corollary}
\newtheorem{rem}{Remark}            

\newtheorem{thm}{Theorem}

\newcommand {\beq}{\begin{equation}}
\newcommand {\eeq}{\end{equation}}
\newcommand {\beqa}{\begin{eqnarray}}
\newcommand {\eeqa}{\end{eqnarray}}         
\newcommand {\beqs}{\begin{eqnarray*}}
\newcommand {\eeqs}{\end{eqnarray*}}
\newcommand {\bds}{\begin{displaymath}}
\newcommand {\eds}{\end{displaymath}}

\newcommand {\n}{\nonumber \\}
\newcommand{\no}{\noindent} 

\newcommand {\eq}[1]{eq.(\ref{#1})}

\newcommand {\Eq}[1]{Eq.(\ref{#1})}

\newcommand {\bebb}{\begin{thebibliography}}
\newcommand {\eebb}{\end{thebibliography}}      
\newcommand {\bbit}{\bibitem}

\def\al{\alpha}
\def\bt{\beta}
\def\Gm{\Gamma}
\def\gm{\gamma}
\def\dl{\delta}

\def\eps{\epsilon}

\def\lm{\lambda}
\def\Lm{\Lambda}

\def\ps{\psi}

\def\sh{sinh}

\def\mpto{\mapsto}

\def\smq{\simeq}

\def\hb2{\frac{\hbar}{2}}

\begin{document}

\baselineskip = 15 pt


\title{
\LARGE\bf Free boson representation of $DY_{\hbar}(gl_N)_k$
}
\vspace{1cm}
\author{
{\normalsize\bf
X.M. Ding $^{a,b}$, B. Y. Hou $^c$, B.Yuan Hou $^d$, L. Zhao $^c$
}\\
\normalsize $^a$ CCAST, P.O. Box 8730, Beijing,100080, China\\
\normalsize $^b$ Institute of Theoretical Physics of Acedimia Sinica,
P.O.Box 2735, 100080, China\\
\normalsize  $^c$ Institute of Modern Physics, Northwest University, Xian, 710069, China\\
\normalsize $^d$ Physics Department of Graduate School of Acedimia Sinica, Beijing, 100039, China\\
}

\date{}

\maketitle

\vspace{2cm}

\begin{abstract}

We construct a realization of the Yangian double $DY_\hbar(gl_N)$ and 
$DY_\hbar(sl_N)$ of an arbitrary level $k$ in terms of free
boson fields with a continuous parameter. In the $\hbar\!\rightarrow\! 0$ 
limit this realization
becomes the Wakimoto realization of $gl_N$ and ${sl_N}$, respectively.
The vertex operators and
the screening currents 
are also constructed with the same spirits; the screening currents commute with $DY_\hbar(sl_N)$ 
modulo total difference. 

\end{abstract}


\section{Introduction}
 
Quantum affine algebras and Yangian were proposed by Drinfeld as 
generalizations
of classical universal Lie algebras and loop algebra with nontrivial Hopf algebra structures
\cite{D1,D2,Dr:new}. They play crucial role, which is similar the role of the   Virasoro algebras and Kac-Moody algebras in the conformal field theories, in the perturbative integrable massive quantum field theories. Following the Faddeev-Reshetikhin-Takhtajan (FRT) formalism
\cite{FRT}, both kinds of
algebras can be considered as associative algebras defined through
the Yang-Baxter relation (i.e. $RLL$-relations) with the structure constants
determined by the solutions of the quantum Yang-Baxter
equation (QYBE). Quantum affine algebra \cite{Dr:new} 
and Yangian double with center (the central extension of the quantum double of  Yangian) \cite{KT,K,KL} are respectively affine extensions of classical  universal Lie algebra and Yangian. From the $RLL$ viewpoint, which corresponds to the trigonometric solution of QYBE and rational one, respectively. If one regards
the Reshetikhin-Semenov-Tian-Shansky realization (RS) \cite{RS} as the affine 
analog of Faddeev-Reshetikhin-Takhtajan formalism. They both were
proved to have important applications in certain physical problems,
especially in describing the dynamical symmetries of the perturbative integrable massive quantum field theories, calculating
the correlation functions and form factors of some two-dimensional
exactly solvable lattice statistical model and (1+1)-dimensional
completely integrable quantum field theories \cite{BL,LS,S2}. 
In such physical applications, the infinite-dimensional representations of quantum affine algebra and Yangian double are required. Rephrasing in another 
words, doubled Yangian with central is needed, especially the representations
with higher ($k>1$) level.

In practice, free field realization is proved to be quite effective and 
useful approach to study complicated algebras structure.
In this aspect, the Heisenberg algebra (or free boson) representation
has become a common method for obtaining representations of
quantum affine algebras. For examples, the free boson representations
of $U_q(\widehat{sl_2})$ with an arbitrary level have been obtained
in Refs.\cite{Srs,M1,M2,kim,abg}.
Free boson representation of $U_q(\widehat{sl_N})$
with level 1 was constructed in \cite{FJ}. Free boson representations
of $U_q(\widehat{sl_3})$ and $U_q(\widehat{sl_N})$ with arbitrary level
were constructed in \cite{sl3} and \cite{sln} respectively.
For the Yangian doubles, the free field representation of $DY_\hbar(sl_2)$ with level $k$ was constructed in \cite{konno}. 
The level 1 free boson representation of
$DY_\hbar(sl_N)$ was given in \cite{iohara}. In Ref.\cite{hzd1} the free field representation of the Yangian double $DY_\hbar(sl_N)$ with arbitrary level $k$ 
are obtained by a simple correspondence between the
quantum affine algebra $U_q(\widehat{sl_N})$ and the Yangian double
$DY_\hbar(sl_N)$ \cite{hzd1}. This correspondence makes our derivation of
free field representation for $DY_\hbar(sl_N)$ greatly simplified.
However, we have been unable to obtain 
screening currents for $DY_\hbar(sl_N)$ following the same spirit, so the relations of quantum affine algebras and the Yangian double with center is nontrivial. So to construct a free fields realization of the screening current operator is the one of the motivations of this paper.

Another motivation is that, in the Ref. \cite{FR} Frenkel and Reshetikhin use  the free field 
realization of the quantum affine algebra $U_q(\widehat{gl_2})_k$ at critical 
level $k=-2$, they investigate the structure of its infinite dimensional center of the quantum affine algebra $U_q(\widehat{gl_2})_k$. Using the Wakimoto realization of the quantum affine algebra they define a new Poisson bracket algebra, which is nothing but 
the $q$-deformation of the classical Virasoro algebra ( in the general case are  the $q$-deformations of the classical ${\cal W}_q (sl_N)$ algebras ). 
Similarly as the quantum affine cases, 
infinite dimensional centers exist for the Yangian double with center at  critical level, which are just the classical 
$\hbar-{\cal W}$ algebras \cite{HY}. In the simplest case, $\hbar$-Virasoro is considered 
in Ref. \cite{DHZ} by using the free fields realization of the Yangian double 
with center $DY_{\hbar}(gl_2)_k$ at level $k=-2$. Similarly, one way to 
study the structure of the 
$\hbar-{\cal W}$ algebras is through the free field realization of the Yangian 
double with center $DY_{\hbar}(g)_k$ at critical level, where the Lie algebra  $g$ has a higher rank. However, it is difficult to obtain the free fields realization of the currents for Yangian double $DY_{\hbar}(gl_N)$ in the higher rank cases (rank greater than one) with the boson fields using in\cite{hzd1}, either no consistent realization of quantum determinant, which decomposes the $DY_{\hbar}(gl_N)_c$ to its 
subalgebra $DY_{\hbar}(gl_N)_c$ could be given.       

In this paper, we introduce free fields involved with a continuous parameter, we obtain the $DY_{\hbar}(gl_N)$ and $DY_{\hbar}(sl_N)$ currents, and  the screening currents 
and vertex operators.

The manuscript is arranged as follows. Section $2$, at first, we briefly review the Drinfeld new realization of the Yangian $Y(g)$, then we give the defining 
relations of Drinfeld generators of the Yangian double with center $DY_{\hbar}(gl_N)$; and the RS realization is also given, the isomorphism 
between these two realization can be established through Ding-Frenkel  correspondence. In section $3$, the three set of free boson fields with a continuous parameter are introduced; some relations, which will be used 
in the following sections are listed. With the help of the results obtained in 
the section $3$, we give a free fields realization of $DY_{\hbar}(gl_N)$ and 
$DY_{\hbar}(sl_N)$ in the section $4$. In section $5$ is the free fields 
realization of the screening currents and the vertex operators, and brief discussions are given at the final section of the paper. 
   
\section{Drinfeld generators of Yangian double $DY_\hbar(gl_N)$}

 At first, we introduce some notations, which will be used in the sequel. Let 
$\eps _{\mu}~~(1 \leq \mu \leq N)$ be the orthonormal basis in $R^N$. We have 
the inner product $<\eps_{\mu} , \eps_{\nu}>=\dl _{\mu ,\nu}$. Define 

\beq
\bar{\eps }_{\mu}=\eps _{\mu} -\eps ,~~~\eps =\frac{1}{N}\sum _{\mu=1} ^N
\eps _{\mu}.
\eeq

\no Obviously, $\sum _{\mu=1} ^N\bar{\eps} _{\mu}=0$. The simple 
roots, fundamental weights are, 

\beq
\al _{\mu}=\eps _{\mu}-\eps _{\mu +1}~,~~~
\Lm _{\mu}=\sum _{\nu=1} ^{\mu}\bar{\eps }_{\nu}~~,
\eeq

\no respectively. For simple, denote half of the Cartan matrix $a_{\mu, \nu}$ as $B_{\mu, \nu}$ .

  There are three realizations of Yangian $Y(g)$: Drinfeld realization 
\cite{D1,D2}, Drinfeld new realization \cite{Dr:new}, and $R$-matrix realization or FRT realization \cite{KR,FRT}. The isomorphism between 
the Drinfeld new realization and the $R$-matrix (or FRT) realization can be established by using the Ding and Frenkel correspondence \cite{DF}. Here, we only give the definition of the Drinfeld new realization of Yangian, for the Yangian double is defined in terms of quantum double of the Drinfeld new realization,  while the Yangian double with center is the central extension of the Yangian double. There is the following theorem: 

\begin{thm}\cite{Dr:new}\label{thm1}
The Yangian $Y(g)$ is isomorphic to the associative algebra with generators, 
$x^{\pm} _{\mu , r},~~ h_{\mu ,r}, ~~\mu=1,\ldots, \mbox{rank}~ g, ~~
r\in Z_{\geq 0}$, and the following defining relations.

\beqs
 & &[ h_{\mu ,r}, h_{\nu ,s}]=0,
\\
 & &[ h_{\mu ,0}, x^{\pm} _{\nu , s}]
=\pm 2d_{\mu}B_{\mu, \nu}x^{\pm} _{\mu , s},
\\
 & &[ h_{\mu ,r+1}, x^{\pm} _{\nu , s}]-[ h_{\mu ,r}, x^{\pm} _{\nu , s+1}]
=\pm \hbar d_{\mu}B_{\mu, \nu}
(h_{\mu , r}x^{\pm} _{\nu , s}+x^{\pm} _{\nu , s}h_{\mu , r}),
\\
& & [ x^+ _{\mu , r}, x^- _{\nu , s}] =\dl _{\mu ,\nu}h_{\mu , r+s},
\\
& & [ x^{\pm} _{\mu , r+1}, x^{\pm} _{\nu , s}]- 
 [ x^{\pm} _{\mu , r}, x^{\pm} _{\nu , s+1}] 
=\pm \hbar d_{\mu}B_{\mu, \nu}
(x^{\pm} _{\mu , r}x^{\pm} _{\nu , s}+ x^{\pm} _{\nu , s}x^{\pm} _{\mu , r}),
\\ 
& &\sum _{\pi} [ x^{\pm} _{\mu, r_ {\pi (1)}}, ~[ x^{\pm} _{\mu, 
r_ {\pi (2)}},\ldots,
 [ x^{\pm} _{\mu, r_{ \pi (m)}},  x^{\pm} _{\nu, s } ]\ldots]]=0,
\eeqs

\no for all sequences of non-negative integers $r_1,\ldots, r_{\nu }$, where 
$m=1-2B_{\mu ,\nu}, ~~2d_{\mu}B_{\mu, \nu}$ is the symmetric Cartan matrix. And the sum is over all permutations $\pi$ of ${1, \ldots, m}$. 
\end{thm}

As an associative algebra, the Yangian double with central $DY_\hbar(gl_N)_c$ is
generated by the Drinfeld generators $\{k^+_{\mu ,r}, k^-_{\mu ,s}~~x^+_{\nu ,m},~~x^-_{\nu ,m} |
\mu=1,~~2,~~...,N; \nu=1,~~2,~~...,N-1;~~r \in Z_{\geq 0},~~s \in Z_{<0},~~
m \in Z \}$, derivation operator $d$ and the center $c$.

The generating function of the Drinfeld generators of Yangian double with center $DY_{\hbar}(gl_N)_c$ are $k^{\pm} _{\mu} (u)$ and  $X^{\pm}_{\nu} (u)$, 

\beqs
& & k^{+}_{\mu} (u) =\sum_{r \geq 0} k_{\mu ,r} u^{-r-1},\;\;
k^{-}_{\mu} (u) =\sum_{s < 0} h_{\nu ,s} u^{-s-1}, \\
& & X^+ _{\mu} (u) = \sum_{m \in Z} x^+_{\mu ,m} u^{-m-1},\;\;
X^-_{\mu} (u) = \sum_{m \in Z} x^-_{\mu ,m} u^{-m-1},
\eeqs

\no and they satisfy the following commutation relations:

\beqs 
& & [ d, \chi(u) ]=\frac{dX}{du},~~~ [ c, \chi (u) ]=0, ~~~ \chi(u)=k^{\pm} _{\mu}(u), ~~X_{\mu} ^+(u),~~X_{\mu} ^-(u) \label{yn0}
\\
& &k_{\mu}^{\pm}(u)k_{\nu}^{\pm}(v)
=k_{\nu}^{\pm}(v)k_{\mu}^{\pm}(u),
\\
& &\rho (u-v-c\hbar)k_{\mu}^{+}(u)k_{\mu}^{-}(v)
=k_{\mu}^{-}(v)k_{\mu}^{+}(u)\rho(u-v),
\\
& &\rho (u-v-c\hbar)\frac{u-v-c\hbar}{u-v+\hbar-c\hbar}k_{\mu}^{+}(u)k_{\nu}^{-}(v)
=k_{\nu}^{-}(v)k_{\mu}^{+}(u)\frac{u-v}{u-v+\hbar}\rho(u-v),~~ \mu <\nu,
\\
& &\rho (u-v-c\hbar)\frac{u-v-\hbar-c\hbar}{u-v-c\hbar}k_{\mu}^{+}(u)k_{\nu}^{-}(v)
=k_{\nu}^{-}(v)k_{\mu}^{+}(u)\frac{u-v-\hbar}{u-v}\rho(u-v),~~\mu >\nu,\\
& &k_{\mu}^{\pm}(u)X_{\mu}^{+}(v)k_{\mu}^{\pm}(u)^{-1}
=\frac{u-v}{u-v+\hbar}X_{\mu}^{+}(v),\\
& &k_{\mu+1}^{\pm}(u)X_{\mu}^{+}(v)k_{\mu+1}^{\pm}(u)^{-1}
=\frac{u-v}{u-v-\hbar}X_{\mu}^{+}(v),\\ 
& &k_{\mu}^{-}(u)X_{\mu}^{-}(v)k_{\mu}^{-}(u)^{-1}
=\frac{u-v+\hbar}{u-v}X_{\mu}^{-}(v),\\
& &k_{\mu +1}^{-}(u)X_{\mu}^{-}(v)k_{\mu +1}^{-}(u)^{-1}
=\frac{u-v-\hbar}{u-v}X_{\mu}^{-}(v),\\
& &k_{\mu}^{+}(u)X_{\mu}^{-}(v)k_{\mu}^{+}(u)^{-1}
=\frac{u-v+\hbar-c\hbar}{u-v-c\hbar}X_{\mu}^{-}(v),\\
& &k_{\mu +1}^{+}(u)X_{\mu}^{-}(v)k_{\mu +1}^{+}(u)^{-1}
=\frac{u-v-\hbar-c\hbar}{u-v-c\hbar}X_{\mu}^{-}(v),\\
& &k_{\mu}^{\pm}(u)X_{\nu}^{+}(v)k_{\mu}^{\pm}(u)^{-1}
=X_{\nu}^{+}(v),~~\mu -\nu\geq 2,~\mbox{or}~ \mu -\nu\leq -1\\
& &k_{\mu}^{\pm}(u)X_{\nu}^{-}(v)k_{\mu}^{\pm}(u)^{-1}
=X_{\nu}^{-}(v), ~~\mu -\nu\geq 2,~\mbox{or}~ \mu -\nu\leq -1,\\            
& &(u-v\mp \hbar)X_{\mu}^{\pm}(u)X_{\mu}^{\pm}(v)
=(u-v\pm \hbar)X_{\mu}^{\pm}(v)X_{\mu}^{\pm}(u),\\
& &(u-v+\hbar)X_{\mu}^+(u)X_{\mu +1}^+(v)
=(u-v)X_{\mu +1}^+(v)X_{\mu}^+(u),\\
& &(u-v)X_{\mu}^-(u)X_{\mu +1}^-(v)
=(u-v+\hbar)X_{\mu +1}^-(v)X_{\mu}^-(u),\\ 
& &X_{\mu}^{\pm}(u_1)X_{\mu}^{\pm}(u_2)X_{\nu}^{\pm}(v)-
              2X_{\mu}^{\pm}(u_1)X_{\nu}^{\pm}(v)X_{\mu}^{\pm}(u_2)\\
             & &~~~+X_{\nu}^{\pm}(v)X_{\mu}^{\pm}(u_1)X_{\mu}^{\pm}(u_2)\\
             & &~~~~~+\{ u_1 \leftrightarrow u_2 \}=0 \quad |\mu -\nu|=1, \\  
& &X_{\mu}^{\pm}(u)X_{\nu}^{\pm}(v)=
               X_{\nu}^{\pm}(v)X_{\mu}^{\pm}(u)\quad |\mu -\nu|\geq 2,\\
& &[X_{\mu}^+(u),X_{\nu}^-(v)]=\delta_{\mu ,\nu}
               \left(\delta(u-(v+c\hbar))
                       k_{\mu +1}^+(u)k_{\mu}^+(u)^{-1}-
                       \delta(u-v)
                       k_{\mu +1}^-(v)k_{\mu}^-(v)^{-1} \right).
\eeqs

\no In which the scalar factor 

\beq
\rho ^{+} (u)=\frac{\Gm(\frac{1}{N}- \frac{u}{N\hbar})
\Gm(1-\frac{1}{N}-\frac{u}{N\hbar})}
{\Gm(- \frac{u}{N\hbar})
\Gm(1- \frac{u}{N\hbar})},
\eeq

\no while $ \delta(u-v)=\sum _{m+n=-1} u^{m}v^{n}$. These relations are understood as relations between formal series. The above relations can be 
derived through the RS realization, in the following we will turn to the 
RS realization.  

The RS realization \cite{RS} originated from the quantum inverse scattering method. The one of the advantages of the RS realization is that, the central 
elements can be written explicit by the quantum trace \cite{RS}. For this purpose, in the following we will drop the $d$ operator in the Yangian double 
with center, and denote as $DY'_{\hbar}(gl_N)$. Introduce the 
following operators:

\beq
L^{\pm}(u)=(l_{\mu \nu }^{\pm}(u))_{1\leq \mu,\nu \leq N};
\eeq

\no where the generating function,

\beq
l_{\mu \nu }^{+}(u)=\dl _{\mu ,\nu} + \hbar\sum _{m=0}^{\infty}
l_{\mu ,\nu}^{+}[m]u^{-m-1}; 
l_{\mu \nu }^{-}(u)=\dl _{\mu ,\nu} - \hbar\sum _{m=-1}^{-\infty}
l_{\mu ,\nu}^{-}[m]u^{-m-1}; 
\eeq

\no and the $R$-matrix $R^{\pm}(u)$, 

\beq
R^{\pm}(u)=\rho ^{\pm}(u)\bar{R}(u),
\eeq

\beq
\rho ^{-} (u)=\left( \frac{\Gm(\frac{1}{N}+\frac{u}{N\hbar})
\Gm(1-\frac{1}{N}+\frac{u}{N\hbar})}
{\Gm(\frac{u}{N\hbar})
\Gm(1+ \frac{u}{N\hbar})}\right)^{-1}=\frac{1}{\rho ^{+}(-u)},
\eeq

\no and $\bar{R}(u)=\frac{uI+\hbar P}{u+\hbar}$ is the Yang's solution of 
QYBE. The $P$ is the flip operator of $u\otimes v$, $Pu\otimes v= v\otimes u$, and the 
choice of the scalar function $\rho ^{\pm} (u)$ can be determined by the unitarity condition and crossing symmetry\cite{IFR}. But here we adopt another approach but the crossing symmetry condition, which will be explained in the latter part of the paper.

From the universal R-matrix for the central extended Yangian double obtained in 
\cite{KT,K,KL}, $L^{\pm}(u)$ have the following unique decompositions:

\beqs
L^{+}(u) =& \left(\matrix{
      1 &0 &\ldots &0\cr
      \hbar f_{2,1}^{+}(u-c\hbar) &1 & &\cr 
           &\vdots &\ddots &\ddots &\cr 
      \hbar f_{N,1}^{+}(u-c\hbar)&\ldots &\hbar f_{N,N-1}^{+}(u-c\hbar) &1
\cr}\right)
 \left(\matrix{
      k_{1}^{+}(u) &0 &\ldots &0\cr
              & \ddots & &\cr  
              & & \ddots &\cr
      0 &\ldots &\ldots & k_{N}^{+}(u)
\cr}\right) \\
& \times \left(\matrix{
            1 & \hbar e_{1,2}^{+}(u)&\ldots & \hbar e_{1,N}^{+}(u)\cr
              & \ddots &\ddots &\cr
              & & \ddots &\hbar  e_{N-1,N}^{+}(u)\cr
            0 &\ldots &\ldots &1
\cr}\right),\\
\label{Lp}
L^{-}(u) &= \left(\matrix{
      1 &0 &\ldots &0\cr
      \hbar f_{2,1}^{-}(u) &1 & &\cr 
           &\vdots &\ddots &\ddots &\cr 
      \hbar f_{N,1}^{-}(u)&\ldots &\hbar f_{N,N-1}^{-}(u) &1
\cr}\right)
 \left(\matrix{
      k_{1}^{-}(u) &0 &\ldots &0\cr
              & \ddots & &\cr  
              & & \ddots &\cr
      0 &\ldots &\ldots & k_{N}^{-}(u)
\cr}\right)\\
& \times \left(\matrix{
            1 & \hbar e_{1,2}^{-}(u)&\ldots & \hbar e_{1,N}^{-}(u)\cr
              & \ddots &\ddots &\cr
              & & \ddots &\hbar  e_{N-1,N}^{-}(u)\cr
            0 &\ldots &\ldots &1
\cr}\right),
\label{Ln}
\eeqs

\no the defining relations in $DY'_{\hbar}(gl_N)$ are given as follows:

\beqa
R^{\pm}(u-v)L^{\pm}_1(u)L^{\pm}_2(v)=
L^{\pm}_2(v)L^{\pm}_1(u)R^{\pm}(u-v), 
\n
R^{+}(u-v-c\hbar)L^{+}_1(u)L^{-}_2(v)=
L^{-}_2(v)L^{+}_1(u)R^{+}
(u-v),
\eeqa

\no and the notation 

\bds
L_1(u)=L(u)\otimes 1,  ~~L_2(u)=1 \otimes L(u),
\eds

\no is used. From the above expression, if we set

\beqs
 &X_{\mu}^-(u)=f_{\mu+1,\mu}^+(u)-f_{\mu+1,\mu}^-(u), \\
 &X_{\mu}^+(u)=e_{\mu,\mu+1}^+(u)-e_{\mu,\mu+1}^-(u).
\eeqs

\no which define an isomorphism between the Drinfeld realization and the RS 
realization. The statement can be proved following the Ding-Frenkel 
equivalence\cite{DF}. 

Similar as $DY_\hbar(gl_N)'_c$, an associative algebra, the Yangian double with central $DY_\hbar(sl_N)'_c$ is a subalgebra of $DY_\hbar(gl_N)'_c$, which 
is obtained with decomposing the $DY_\hbar(gl_N)'_c$ by the quantum 
determinant. If we define the following currents:

\beqa
& &H_{\mu} ^{\pm}(u)=k_{\mu +1}^{\pm}(u+\frac{\mu}{2}\hbar)
              k_{\mu}^{\pm}(u+\frac{\mu}{2}\hbar )^{-1}~,\n
& &K^{\pm}(u)=\prod_{\mu=1}^N 
   k_{\mu} ^{\pm}\left( u+(\mu -1)\hbar\right), \n
& & E_{\mu} (u)=\frac{1}{\hbar}X_{\mu} ^{+}(u+\frac{\mu}{2}\hbar)~,~
   F_{\mu} (u)=\frac{1}{\hbar}X_{\mu} ^{-}(u+\frac{\mu}{2}\hbar). 
\label{decom}
\eeqa

\no and their modes expansion are,

\beqs
& &H^{+}_{\mu}(u) = 1 + \hbar \sum_{m \geq 0} h_{\mu ,m} u^{-m-1},\;\;
H^{-}_{\mu}(u) = 1 - \hbar \sum_{m < 0} h_{\mu ,m} u^{-m-1},
\\
& &E _{\mu}(u) = \sum_{m \in Z} e_{\mu ,m} u^{-m-1},\;\;
F_{\mu}(u) = \sum_{m \in Z} f_{\mu ,m} u^{-m-1}.
\eeqs

\no The Drinfeld generators of $DY'_\hbar(sl_N)_c$ are  $\{h_{\nu ,m}, 
~~e_{\nu ,m},~~f_{\nu ,m},~~c |
\nu=1,~~2,~~...,N-1;~~m \in Z  \}$. 

\no Now we can write the generating relations for $DY'_\hbar(sl_N)$
as follows \cite{iohara},

\beqa
& & [ H_{\mu}^\pm(u),~H_{\nu}^\pm(v) ] =0,~~~[c,\mbox{everything}]=0~,
\label{yn1}\\
& & (u-v+ B_{\mu ,\nu} \hbar\mp c\hbar) (u-v - B_{\mu ,\nu} \hbar)
H_{\mu}^{\pm}(u) H_{\nu}^{\mp}(v) \n
& &~~~~= (u-v- B_{\mu ,\nu} \hbar\mp c\hbar) (u-v + B_{\mu ,\nu} \hbar)
H_{\nu}^{\mp}(v) H_{\mu}^{\pm}(u), \label{yn2} \\
& & (u-v - B_{\mu ,\nu} \hbar) H_{\mu}^{\pm}(u) E_{\nu}(v)
= (u-v + B_{\mu ,\nu} \hbar) E_{\nu}(v) H_{\mu}^{\pm}(u), \label{yn3} \\
& & (u-v + B_{\mu ,\nu} \hbar-c\hbar) H_{\mu}^+(u) F_{\nu}(v)
= (u-v - B_{\mu ,\nu} \hbar-c\hbar) F_{\nu}(v) H_{\mu}^+(u), \label{yn4} \\
& & (u-v + B_{\mu ,\nu} \hbar) H_{\mu}^-(u) F_{\nu}(v)
= (u-v - B_{\mu ,\nu} \hbar) F_{\nu}(v) H_{\mu}^-(u), \label{yn4F} \\
& & (u-v - B_{\mu ,\nu} \hbar) E_{\mu}(u) E_{\nu}(v)
=  (u-v + B_{\mu ,\nu} \hbar) E_{\nu}(v) E_{\mu}(u), \label{yn5} \\
& & (u-v + B_{\mu ,\nu} \hbar) F_{\mu}(u) F_{\nu}(v)
=  (u-v - B_{\mu ,\nu} \hbar) F_{\nu}(v) F_{\mu}(u), \label{yn5F} \\
& & [ E_{\mu}(u),~F_{\nu}(v) ] =
\frac{\dl _{\mu ,\nu}}{\hbar} \left(
\delta( u - (v+c\hbar)) H^+_{\mu}(u) - \delta( u - v) H^-_{\mu}(v) \right),
\label{yn6} \\
& & E_{\mu}(u_1) E_{\mu}(u_2) E_{\nu}(v)
-2 E_{\mu}(u_1) E_{\nu}(v) E_{\mu}(u_2) \n
& &~~~~~+ E_{\nu}(v) E_{\mu}(u_1) E_{\mu}(u_2) +
(\mbox{replacement:}~u_1 \leftrightarrow u_2) = 0
~\mbox{for}~|\mu -\nu|=1, \label{yn7}\\
& & F_{\mu}(u_1) F_{\mu}(u_2) F_{\nu}(v)
-2 F_{\mu}(u_1) F_{\nu}(v) F_{\mu}(u_2) \n
& &~~~~~+ F_{\nu}(v) F_{\mu}(u_1) F_{\mu}(u_2) +
(\mbox{replacement:}~u_1 \leftrightarrow u_2) = 0
~\mbox{for}~|\mu- \nu|=1, \label{yn7F}\\
& & E_{\mu}(u) E_{\nu}(v)
= E_{\nu}(v) E_{\mu}(u) ~~\mbox{for}~|\mu- \nu|>1, \label{yn8}\\
& & F_{\mu}(u) F_{\nu}(v)
= F_{\nu}(v) F_{\mu}(u) ~~\mbox{for}~|\mu- \nu|>1, \label{yn8F}\\
& & [K^{\pm}(u),K^{\pm}(v)]=0, \label{yn9}\\
& & K^{+}(u)K^{-}(v)=K^{-}(v)K^{+}(u),\label{yn10}\\
& & [K^{\eps}(u),H_{\mu}^{\pm}(v)]=[K^{\eps}(u),E_{\mu}(v)]\n
& &~~ =[K^{\eps}(u),F_{\mu}(v)]=0,~~\forall \eps=\pm, \forall 1\leq \mu\leq N. 
\eeqa

\no Which are obtained by direct calculation. From the above commuting relations, $ K^{\pm}(u) $  
commute with all of the elements of $DY_{\hbar}(gl_N)'_c $, so the Fourier coefficients of the power series $K(u)-1$ are the central elements of the $DY_{\hbar}(gl_N)'_c$. It is obvious that the Heisenberg subalgebra of $DY_{\hbar}(gl_N)'_c$ is generated by the quotient of $DY_{\hbar}(gl_N)'_c$ 
by the ideal generated by this central elements. In fact we see that the formula $K^{\pm}(u)=q-det.L^{\pm}(u)$, and the scalar function is chosen so that the quantum determinant commuting with the all of the currents and with itself.

\section{Free boson with continuous parameter}

In this section we shall construct a free boson representation of $DY_\hbar(sl_N)_k$ with arbitrary level $k$ ($c=k$ henceforth).
For $N=2$ this problem has already been solved in Ref.
\cite{konno}. For generic $N$, the Yangian double currents are obtained using the quantum-Yangian double correspondence \cite{hzd1}, but however, screening currents and vertex operators can not be get in the same way, at least, they 
can not be expressed as a nice form. So the  correspondence between them are somewhat nontrivial. In the next section, we introduce another kinds 
Heisenberg algebras to overcome these problems.

$N^2 -1$ free boson fields are needed to construct the $DY_{\hbar}(sl_N)$ generators with 
arbitrary level. So we introduce three set boson fields $\hat{\lm }_{\mu}\;
(1\leq \mu \leq
N)$, $\hat{b}_{\mu ,\nu}$ and $\hat{c}_{\mu ,\nu}\; (1\leq \mu < \nu \leq N)$, and while 
$\hat{b}_{\mu ,\nu}=\hat{c}_{\mu ,\nu}\equiv 0$, for $\mu\geq \nu$. 
The quantum Heisenberg algebra ${\cal A}_{\hbar , k}(sl_N)$ generated by 
$\hat{\lm }_{\mu}(t),~~\hat{b}_{\mu ,\nu}(t),~~\hat{c}_{\mu ,\nu}(t) $
is involved with continuous parameters $t\;(t\in R-0)$, which enjoy the the following commutation relations 

\beqa
& & [ \hat{\lm} _{\mu }(t),~\hat{\lm} _{\mu } (t') ] 
=\frac{\sh\frac{k+g}{2}\hbar t 
\sh \frac{(N-1)}{2}\hbar t\sh\hb2 t }{\hbar ^2 \sh \frac{N}{2}\hbar t} \delta(t+t'),
\label{bs1}\\ 
& & [ \hat{\lm} _{\mu }(t),~\hat{\lm} _{\nu } (t') ] =-\frac{\sh\frac{k+g}{2}\hbar t 
\sh ^2\hb2 t }{\hbar ^2 \sh \frac{N}{2}\hbar t}e^{sign(\mu -\nu)\frac{N}{2}t} \delta(t+t'),\\
& &\sum _{\mu =1} ^N e^{(1-\mu)\hbar t}\hat{\lm} _{\mu}(t)=0;\\
& & [ \hat{b}_{\mu ,\nu }(t), \hat{b}_{\mu ' ,\nu '}(t') ] = -\frac{1}{\hbar ^2 t}\sh ^2\frac{\hbar t}{2}
\delta _{\mu  ,\mu  '} \delta _{\nu  ,\nu  '}\delta (t+t'), 
\label{bs2}\\
& &  [ \hat{c}_{\mu ,\nu }(t), \hat{c}_{\mu ',\nu'}(t') ] =\frac{1}{\hbar ^2 t}\sh ^2\frac{\hbar t}{2}
\delta_{\mu  ,\mu  '} \delta_{\nu  ,\nu  '}\delta (t+t'), 
\label{bs3}
\eeqa

\no and the other commutators vanish identical, where $g=N$ is the dual Coxeter number for the $sl_N$ algebra. Here $\hat{\lm }_{\mu}(t)$ are the 
"fundamental weight type" fields, while  $\hat{b}_{\mu ,\nu}(t),
~~\hat{c}_{\mu ,\nu}(t) $ are "ghost type" fields in conformal field theory 
alike. In fact, the free boson fields defined above 
can be viewed as the Yangian limit of the elliptic algebra 
${\cal A}_{\hbar ,\eta} (\widehat{sl_N})$, with $\eta \rightarrow 0$  \cite{KLP}.

The Fock space corresponding to the above Heisenberg algebras can be
specified as follows. Let $| 0 >$ be the vacuum state defined by

\beq
\lm _{\mu } (t)| 0 > = b_{\mu ,\nu }(t)| 0 >= c_{\mu ,\nu }(t)| 0 >
=0~,~~~ ( t>0),
\eeq

\noindent The Fock space ${\cal F}(l_{\lm},l_b,l_c)$ is then generated
by the actions of the negative modes of $\lm   _{\mu},~b_{\mu ,\nu},~c_{\mu ,\nu}$. We shall
see later that this Fock space actually forms a (Wakimoto-like \cite{wakimoto,fff})
module for the Yangian double $DY_\hbar(sl_N)$ with level $k$.

We define the free boson fields $\chi _{\mu ,\nu } (\beta|u;\alpha)$ and  
$\chi _{lm} ^{\pm} (u)$ through

\beqa
 & &\chi _{\mu ,\nu }(\beta|u;\alpha)=exp\{-\hbar \int _{-\infty} ^0\frac{dt}{2\pi i}
\frac{1}
{\sh\frac{\beta}{2}\hbar t}\hat{\chi} _{\mu ,\nu } (t)e^{\alpha \hbar t}e ^{-iut}
\}\n
& &~~~~~~~~~~~~~~~~~~exp\{-\hbar \int _0 ^{+\infty}\frac{dt}{2\pi i}
\frac{1}{\sh\frac{\beta}{2}\hbar t}
\hat{\chi} _{\mu ,\nu } (t)e^{-\alpha \hbar t}e ^{-iut}\},\\
 & & \chi _{\mu ,\nu } ^{+} (u) =exp\{2\hbar  \int _0 ^{+\infty}\frac{dt}{2\pi i} \hat{\chi} _{\mu ,\nu } (t)e ^{-iut}\}
\\
& & \chi _{\mu ,\nu } ^{-} (u)=exp\{-2\hbar \int _{-\infty} ^0\frac{dt}{2\pi i} 
\hat{\chi} _{\mu ,\nu } (t)e ^{-iut}\},
\eeqa

\no where $\chi _{\mu ,\nu }(u)$ ~( $\chi _{\mu ,\nu } ^{\pm}(u)$) stands for 
$\lm _{\nu} (u),\; b_{\mu ,\nu }(u) $ or 
$c_{\mu ,\nu } (u)$ ~( $\lm _{\nu} ^{\pm}(u),\; b_{\mu ,\nu } ^{\pm}(u) $ or 
$c_{\mu ,\nu } ^{\pm} (u)$). Then following the standard procedure we have 
\footnote{Here and below, all OPE relations should be understood
to hold in the analytic continuation sense.}

\beqa
b_{\mu ,\nu }(u)b_{\mu ' ,\nu '} ^{-}(v)& &=exp\left\{-2\int _{\tilde{c}}
\frac{dt \ln(-t)}{2\pi i} \frac{\sh \hb2 t }{t}\dl _{\mu  ,\mu  '}\dl _{\nu  ,\nu  '}
e^{-i(u-v)t} \right\}:b_{\mu ,\nu }(u)b_{\mu ' ,\nu '}^-(v) : \n
& &=\left (
 \frac{\Gm ^2 (\frac{i(u-v)}{\hbar}+\frac{1}{2})}
{\Gm (\frac{i(u-v)}{\hbar}-\frac{1}{2})
 \Gm (\frac{i(u-v)}{\hbar}+\frac{3}{2})}
\right) ^{\dl _{\mu ,\mu '} \dl _{\nu  ,\nu  '}}b_{\mu ' ,\nu '}^- (v)b_{\mu ,\nu }(u)\n
& &=\left ( 
\frac {u-v+i\hb2}{u-v-i\hb2}
\right ) ^{\dl _{\mu  ,\mu  '} \dl _{\nu  ,\nu  '}}b_{\mu ' ,\nu '}^-(v)b^{l,m}(u) 
\label{bb-}
\eeqa

\no where the $\tilde{c}$ indicate the circle  from $+\infty$ to $0$ in the 
upper half plain and $0$ to $\infty$ in the lower half plain \cite{JM,KLP}. To 
derive the \eq{bb-}, the identity

\beq
\int _{\tilde{c}} \frac{dt \ln(-t)}{2i\pi t}\frac{e^{-ut}}{1-e^{-t/ \eta}}
=\ln \Gm(\eta u)+(\eta u -1/2)(\gamma -\ln \eta) -\frac{1}{2}\ln 2\pi,
\eeq

\no is used, in which the $\gamma$ is the Euler number. In the same way, 
we can get the following identities.

\beqa
& & b_{\mu ,\nu }^+(u)b_{\mu ' ,\nu '} (v)=\left ( \frac {u-v+i\hb2}{u-v-i\hb2}\right) ^{\dl _{\mu  ,\mu  '} \dl _{\nu  ,\nu  '}}b_{\mu ' ,\nu '}(v)b_{\mu ,\nu }^+(u), 
\label{b+b}\\
& &\frac{c_{\mu ,\nu }(u) c_{\mu ' ,\nu '} ^{-}(v)}{:c_{\mu ' ,\nu '}^{-}(v)c_{\mu ,\nu }(u):}
=\left ( \frac {u-v-i\hb2}{u-v+i\hb2}\right ) ^{\dl _{\mu  ,\mu  '} \dl _{\nu  ,\nu  '}}, \\
& &\frac{c_{\mu ,\nu }^{+}(u)c_{\mu ' ,\nu '}(v)}{:c_{\mu ' ,\nu '}(v)c_{lm} ^{+}(u):}
=\left ( \frac {u-v-i\hb2}{u-v+i\hb2}\right ) ^{\dl _{\mu  ,\mu  '} \dl _{\nu  ,\nu  '}},
\label{}
\\
& &\frac{\lm_{\mu}^{+}(u) \lm _{\mu}(k+N|v;\alpha) }
        {:\lm _{\mu}(k+N|v;\alpha) \lm _{\mu} ^{+}(u):}
=\frac{\lm _{\mu}(k+N|u;\alpha)\lm _{\mu}^{-}(v)}
{:\lm _{\mu} ^{-}(v)\lm _{\mu}(k+N|u;\alpha):}\n
& &=\frac{\Gm(\frac{i(u-v)}{N\hbar}+\frac{\alpha}{N} )
      \Gm(\frac{i(u-v)}{N\hbar}+\frac{\alpha}{N}+1)}
     {\Gm(\frac{i(u-v)}{N\hbar}+\frac{\alpha +1}{N})
      \Gm(\frac{i(u-v)}{N\hbar}+\frac{\alpha -1}{N} +1)},\\
& &\frac{\lm_{\mu}^{+}(u) \lm _{\nu}(k+N|v;\alpha) }
        {:\lm _{\nu}(k+N|v;\alpha) \lm _{\nu} ^{+}(u):}
=\frac{\lm _{\mu}(k+N|u;\alpha)\lm _{\nu}^{-}(v)}
{:\lm _{\nu } ^{-}(v)\lm _{\mu}(k+N|u;\alpha):}\n
& &=\frac{{\Gm}^2(\frac{i(u-v)}{N\hbar}+\frac{\alpha}{N} )}
     {\Gm(\frac{i(u-v)}{N\hbar}+\frac{\alpha  +1}{N})
      \Gm(\frac{i(u-v)}{N\hbar}+\frac{\alpha -1}{N})}~~~~\mu >\nu,\\
& &\frac{\lm_{\mu}^{+}(u) \lm _{\nu}(k+N|v;\alpha) }
        {:\lm _{\nu}(k+N|v;\alpha) \lm _{\mu} ^{+}(u):}
=\frac{\lm _{\mu}(k+N|u;\alpha)\lm _{\nu}^{-}(v)}
{:\lm _{\nu } ^{-}(v)\lm _{\mu}(k+N|u;\alpha):}\n
& &=\frac{{\Gm}^2(\frac{i(u-v)}{N\hbar}+\frac{\alpha}{N} +1 )}
     {\Gm(\frac{i(u-v)}{N\hbar}+\frac{\alpha  +1}{N}+1)
      \Gm(\frac{i(u-v)}{N\hbar}+\frac{\alpha -1}{N} +1)} ~~~\mu < \nu,
\eeqa

\no and 

\beqa
\lm _{\mu} ^+(u)\lm _{\mu} ^-(v)
&=&\frac{\Gm(\frac{i(u-v)}{N\hbar}+1-\frac{1}{N} -\frac{k+N}{2N} )
      \Gm(\frac{i(u-v)}{N\hbar}+\frac{k+N}{2N} )}
      {\Gm(\frac{i(u-v)}{N\hbar}+1-\frac{1}{N}+\frac{k+N}{2N})
      \Gm(\frac{i(u-v)}{N\hbar}-\frac{k+N}{2N})}\n
& &\times
\frac{\Gm(\frac{i(u-v)}{N\hbar}+\frac{1}{N}-\frac{k+N}{2N} )
      \Gm(\frac{i(u-v)}{N\hbar}+1+\frac{k+N}{2N} )}
     {\Gm(\frac{i(u-v)}{N\hbar}+\frac{1}{N}+\frac{k+N}{2N} )
      \Gm(\frac{i(u-v)}{N\hbar}+1-\frac{k+N}{2N} )}
\lm _{\mu} ^-(v)\lm _{\mu} ^+(u),\n
&=&\frac{\rho^{+}(i(u-v)-\frac{k+N}{2}\hbar)}
{\rho^{+}(i(u-v)+\frac{k+N}{2}\hbar)}
\lm _{\mu} ^-(v)\lm _{\mu} ^+(u),\n
\lm _{\mu} ^+(u)\lm _{\nu } ^-(v)
&=&\frac{\Gm(\frac{i(u-v)}{N\hbar}-\frac{1}{N} -\frac{k+N}{2N} )
      \Gm(\frac{i(u-v)}{N\hbar}+\frac{1}{N}-\frac{k+N}{2N} )}
     {\Gm(\frac{i(u-v)}{N\hbar}-\frac{1}{N}+\frac{k+N}{2N})
     \Gm(\frac{i(u-v)}{N\hbar}+\frac{1}{N}+\frac{k+N}{2N} )}\n
 & &\times
\frac{{\Gm}^2(\frac{i(u-v)}{N\hbar}+\frac{k+N}{2N})}
{{\Gm}^2(\frac{i(u-v)}{N\hbar}-\frac{k+N}{2N})}
\lm _{\nu } ^-(v)\lm _{\mu} ^+(u)~~~~\mu > \nu,\\
\lm _{\mu} ^+(u)\lm _{\nu } ^-(v)
&=&\frac{\Gm(\frac{i(u-v)}{N\hbar}+1-\frac{1}{N} -\frac{k+N}{2N} )
      \Gm(\frac{i(u-v)}{N\hbar}+1+\frac{1}{N}-\frac{k+N}{2N} )}
     {\Gm(\frac{i(u-v)}{N\hbar}+1-\frac{1}{N}+\frac{k+N}{2N})
     \Gm(\frac{i(u-v)}{N\hbar}+1+\frac{1}{N}+\frac{k+N}{2N} )}\n
 & &\times
\frac{{\Gm}^2(\frac{i(u-v)}{N\hbar}+1+\frac{k+N}{2N} )}
{{\Gm}^2(\frac{i(u-v)}{N\hbar}+1-\frac{k+N}{2N})}
\lm _{\nu } ^-(v)\lm _{\mu} ^+(u)~~~~\mu < \nu,\\
b_{\mu ,\nu }^+(u)b_{\mu ' ,\nu '} ^{-}(v)
&=&
\left(\frac{\Gm ^3(\frac{i(u-v)}{\hbar})\Gm (\frac{i(u-v)}{\hbar}+2)}
{\Gm ^3(\frac{i(u-v)}{\hbar}+1)\Gm (\frac{i(u-v)}{\hbar}-1)}
\right) ^{\dl _{\mu  ,\mu  '} \dl _{\nu  ,\nu  '}}b_{\mu ' ,\nu '}^-(v)b_{\mu ,\nu } ^+(u) \n
&=&\left ( \frac {(u-v)^2}{(u-v-i\hbar)(u-v+i\hbar)}\right ) ^{\dl _{\mu  ,\mu  '} \dl _{\nu  ,\nu  '}}b_{\mu ' ,\nu '}^-(v)b_{\mu ,\nu } ^+(u),\\ 
c_{\mu ,\nu }^+(u)c_{\mu ' ,\nu '} ^{-}(v)
&=&
\left(\frac{\Gm ^3(\frac{i(u-v)}{\hbar}+1)\Gm (\frac{i(u-v)}{\hbar}-1)}
{\Gm ^3(\frac{i(u-v)}{\hbar})\Gm (\frac{i(u-v)}{\hbar}+2)}
\right) ^{\dl _{\mu  ,\mu  '} \dl _{\nu  ,\nu  '}}c_{\mu ' ,\nu '}^-(v)c_{\mu ,\nu } ^+(u) \n
&=&\left ( \frac {(u-v-i\hbar)(u-v+i\hbar)}{(u-v)^2}\right ) ^{\dl _{\mu  ,\mu  '} \dl _{\nu  ,\nu  '}}c_{\mu ' ,\nu '}^-(v)c_{\mu ,\nu } ^+(u).
\eeqa 

\no In the process of constructing the boson realization of 
$DY_{\hbar}(sl_N)_k$, the following four identities are useful. In fact, the relations of the arguments in different terms are decided by these identities.  There are 

\beqa
& &\frac{b_{\mu ,\nu} ^+(u):b_{\mu ',\nu '}^-(v)b_{\mu ',\nu '}(v-i\hb2):}
{b_{\mu,\nu} ^+(u)b_{\mu ',\nu '}(v+i\hb2)}=1,\\
& &\frac{b_{\mu,\nu} ^+(u):b_{\mu',\nu'}^-(v+i\hb2)b_{\mu',\nu'}(v+i\hbar)^{-1}:}
{b_{\mu,\nu} ^+(u)b_{\mu ',\nu '}(v)^{-1}}=1,\\
& &\frac{:b_{\mu,\nu} ^+(u+i\hb2)b_{\mu,\nu}^-(u)^{-1}:b_{\mu ',\nu '}^-(v)}
{b_{\mu,\nu} ^+(u+i\hbar)^{-1}b_{\mu ',\nu '}^-(v)}=1,\\
& &\frac{:b_{\mu,\nu} ^+(u)b_{\mu, \nu }^-(u+i\hb2):b_{\mu ',\nu '}^-(v):}
{b_{\mu,\nu} ^+(u-i\hb2)b_{\mu ',\nu'}^-(v)}=1,
\eeqa

\section{Free boson representation of $DY_\hbar(gl_N)_k$ and 
$DY_\hbar(sl_N)_k$ }
In this section, by using the results obtained in the above sections, we 
can define a $DY_{\hbar}(sl_N)_k$ module on the ${\cal F}_{\lm ,~b ,~c}$. Then 
with the help of \Eq{decom},  we get a module of $DY_{\hbar}(sl_N)_k$ on the  ${\cal F}_{\lm ,~b ,~c}$. The results can be expressed in terms as:

\begin{prop}\label{prop1}
There is a homomorphism ${\ps}_{\hbar ,k}$ from $DY_{\hbar}(sl_N)_k $ to 
${\cal A}_{\hbar ,k}(sl_N)$, which is defined on the generators as follows:

\beqa
k^{+}  _{\mu}(u)&\mpto& : \lm   _{\mu} ^{+}(u +\frac{k-N}{4}i\hbar)^{-1}   
\prod_{\al=1} ^{\mu-1} \prod _{\bt=\mu+1}^N 
\frac{b_{\al,\mu} ^{+}(u +\frac{\mu-\al}{2}i\hbar)}
{b_{\mu,\bt}^{+}(u+\frac{\mu-\bt}{2}i\hbar)},
\label{dkp}
\\
k^{-}  _{\mu}(u)&\mpto& : \lm _{\mu } ^{-}(u -\frac{k-N}{4}i\hbar)^{-1}   
\prod_{\rho=1}^{\mu-1} \prod _{\al=1}^{\mu}
\prod_{\gm =\mu}^{N} \prod _{\bt=\mu+1}^N  
\frac{b_{\rho,\gm} ^{-}(u+\frac{\rho +\gm}{2}i\hbar)}
{b_{\al,\bt} ^{-}(u +\frac{\al +\bt -2}{2}i\hbar)},\n
&    & ~~~~~~~~~~~~~~~~~~~~~~~~~~1\leq \mu\leq N;\label{dkn}\\
X^{+}_{\nu }(u) &\mpto& \sum_{m=1}^{\nu}:(b+c)_{m,\nu}(u-\frac{m-\nu-1}{2}i\hbar)
\left\{\frac{b_{m,\nu +1}^{+}(u -\frac{m-\nu-1}{2}i\hbar)}
{(c+b)_{m,\nu +1}(u -\frac{m-\nu}{2}i\hbar)}\right.\n
& &~~~-\left.\frac{b_{m,\nu +1}^{-}(u-\frac{m-\nu-1}{2}i\hbar)}
{(c+b)_{m,\nu +1}(u -\frac{m-\nu-2}{2}i\hbar)}\right\}
\prod _{n=1}^{m-1} 
\frac{b_{n,\nu +1}^{+}(u -\frac{n-\nu-1}{2}i\hbar)}
{b_{n,\nu }^{+}(u -\frac{n-\nu}{2}i\hbar)}:, 
\label{dXX}\\
X^{-}_{\nu }(u)&\mpto&:\left\{
\sum_{m=1}^{\nu} (b+c)_{m,\nu +1}(u + \frac{\nu +m}{2}i \hbar ) 
 \lm _{\nu } ^{-}(u -\frac{k-N}{4}i\hbar)  
\lm _{\nu+1} ^{-}(u -\frac{k-N}{4}i\hbar)^{-1} \right. \n
& &~~~~\times \left[(1+\dl _{m,\nu})
 b_{m,\nu}^- ( u + \frac{\nu +m}{2} i\hbar ) ^{-1}
(b+c)_{m,\nu} ( u +\frac{\nu +m-1}{2} i \hbar )^{-1}\right.\n
& &~~~~ -\left. b_{m,\nu}^+ ( u +\frac{\nu +m}{2} i\hbar )^{-1}
(b+c)_{m,\nu} ( u + \frac{\nu +m+1}{2} i\hbar )^{-1} \right] \n
& &~~~~\times \prod_{n=m+1}^{\nu}\prod_{s=\nu+1}^N
\frac { b_{n,\nu +1}^{-}( u +\frac{\nu +n-1}{2}i \hbar)}
{ b_{n,\nu }^{-}( u +\frac{\nu +n}{2}i \hbar)}
\frac { b_{\nu,s+1}^{-}( u +\frac{\nu +s}{2}i \hbar)}
{ b_{\nu+1,s}^{-}( u +\frac{\nu +s-1}{2}i \hbar)}\n
& &~~~~-\sum_{m=\nu+1}^{N} (b+c)_{\nu ,m}(u - \frac{2k+m-\nu-1}{2}i \hbar )\n 
& &~~~~~~\times \lm _{\nu } ^{+}(u -\frac{3k+g}{4}i\hbar)  
\lm _{\nu+1} ^{+}(u -\frac{3k+g}{4}i\hbar)^{-1}\n 
& &~~~~\times \left[
 (1+\dl _{m,\nu +1})\frac{b_{\nu+1,m}^+ ( u - \frac{2k+m-\nu-1}{2} i\hbar )}
{(b+c)_{\nu+1,m} ( u -\frac{2k+m-\nu}{2} i \hbar )}\right.\n
 & &~~~~~ \left.-\ \frac{b_{\nu+1,m}^- ( u - \frac{2K+m-\nu-1}{2} i\hbar )}
{(b+c)_{\nu+1,m} ( u - \frac{2k+m-\nu-2}{2} i\hbar )} \right]\n
& &\left.~~~~\times \prod_{n=m}^N
\frac { b_{\nu,n}^{+}( u -\frac{2k+n-\nu}{2}i \hbar)}
{ b_{\nu+1,n}^{+}( u -\frac{2k+n-\nu-1}{2}i \hbar)}\right\}:,
~~~1\leq \nu \leq N-1; 
\label{dXY}
\eeqa

\no is well defined ${\cal F}_{\lm ,~b ,~c}$, and satisfies the commutation 
relations:  

\beqs
& &k_{\mu}^{\pm}(u)k_{\nu}^{\pm}(v)
= k_{\nu}^{\pm}(v)k_{\mu}^{\pm}(u),
\\
& &\rho (u-v+ik\hbar)k_{\mu}^{+}(u)k_{\mu}^{-}(v)
=k_{\mu}^{-}(v)k_{\mu}^{+}(u)\rho(u-v),
\\
& &\rho (u-v+ik\hbar)\frac{u-v+ik\hbar}{u-v-i\hbar+ik\hbar}k_{\mu}^{+}(u)k_{\nu}^{-}(v)
=k_{\nu}^{-}(v)k_{\mu}^{+}(u)\frac{u-v}{u-v-i\hbar}\rho(u-v),~~ \mu <\nu,
\\
& &\rho (u-v+ik\hbar)\frac{u-v+i\hbar+ik\hbar}{u-v+ik\hbar}k_{\mu}^{+}(u)k_{\nu}^{-}(v)
=k_{\nu}^{-}(v)k_{\mu}^{+}(u)\frac{u-v+i\hbar}{u-v}\rho(u-v),~~\mu >\nu,\\
& &k_{\mu}^{\pm}(u)X_{\mu}^{+}(v)k_{\mu}^{\pm}(u)^{-1}
=\frac{u-v}{u-v-i\hbar}X_{\mu}^{+}(v),\\
& &k_{\mu+1}^{\pm}(u)X_{\mu}^{+}(v)k_{\mu+1}^{\pm}(u)^{-1}
=\frac{u-v}{u-v+i\hbar}X_{\mu}^{+}(v),\\ 
& &k_{\mu}^{-}(u)X_{\mu}^{-}(v)k_{\mu}^{-}(u)^{-1}
=\frac{u-v-i\hbar}{u-v}X_{\mu}^{-}(v),\\
& &k_{\mu +1}^{-}(u)X_{\mu}^{-}(v)k_{\mu +1}^{-}(u)^{-1}
=\frac{u-v+i\hbar}{u-v}X_{\mu}^{-}(v),\\
& &k_{\mu}^{+}(u)X_{\mu}^{-}(v)k_{\mu}^{+}(u)^{-1}
=\frac{u-v-i\hbar+ik\hbar}{u-v+ik\hbar}X_{\mu}^{-}(v),\\
& &k_{\mu +1}^{+}(u)X_{\mu}^{-}(v)k_{\mu +1}^{+}(u)^{-1}
=\frac{u-v+i\hbar+ik\hbar}{u-v+ik\hbar}X_{\mu}^{-}(v),\\
& &k_{\mu}^{\pm}(u)X_{\nu}^{+}(v)k_{\mu}^{\pm}(u)^{-1}
=X_{\nu}^{+}(v),~~\mu -\nu\geq 2,~\mbox{or}~ \mu -\nu\leq -1\\
& &k_{\mu}^{\pm}(u)X_{\nu}^{-}(v)k_{\mu}^{\pm}(u)^{-1}
=X_{\nu}^{-}(v), ~~\mu -\nu\geq 2,~\mbox{or}~ \mu -\nu\leq -1,\\            
& &(u-v\pm i\hbar)X_{\mu}^{\pm}(u)X_{\mu}^{\pm}(v)
\smq (u-v\mp i\hbar)X_{\mu}^{\pm}(v)X_{\mu}^{\pm}(u)\sim reg.,\\
& &(u-v-i\hbar)X_{\mu}^+(u)X_{\mu +1}^+(v)
\smq (u-v)X_{\mu +1}^+(v)X_{\mu}^+(u)\sim reg.,\\
& &(u-v)X_{\mu}^-(u)X_{\mu +1}^-(v)
\smq (u-v-i\hbar)X_{\mu +1}^-(v)X_{\mu}^-(u)\sim reg.,\\ 
& &X_{\mu}^{\pm}(u_1)X_{\mu}^{\pm}(u_2)X_{\nu}^{\pm}(v)-
              2X_{\mu}^{\pm}(u_1)X_{\nu}^{\pm}(v)X_{\mu}^{\pm}(u_2)\\
             & &~~~+X_{\nu}^{\pm}(v)X_{\mu}^{\pm}(u_1)X_{\mu}^{\pm}(u_2)\\
             & &~~~~~+\{ u_1 \leftrightarrow u_2 \}\sim 0 \quad |\mu -\nu|=1, \\  
& &X_{\mu}^{\pm}(u)X_{\nu}^{\pm}(v)\smq 
               X_{\nu}^{\pm}(v)X_{\mu}^{\pm}(u)\sim reg.
\quad |\mu -\nu|\geq 2,\\
& &[X_{\mu}^+(u),X_{\nu}^-(v)]\smq reg. +\delta_{\mu ,\nu}
               \left(\delta(u-(v-ik\hbar))
                       k_{\mu +1}^+(u)k_{\mu}^+(u)^{-1}\right.\\
& &\left.~~~~~~~~~~~~~~~~~~~~~~-\delta(u-v)k_{\mu +1}^-(v)k_{\mu}^-(v)^{-1} \right).
\eeqs

\no where $reg.$ means some regular terms and $\simeq$
and $\sim$ imply ``equals up to' with difference of such terms.

\end{prop}

\no {\it Proof}: A somewhat long but straightforward calculation verifies this proposition. Using the \Eq{decom} and \Eq{dkp} to \Eq{dXY}, we have the following  expressions: 

\beqa
H^{+}_{\mu }(u)&\mpto& : \lm _{\mu } ^{+}(u +\frac{k-N}{4}i\hbar-i\mu\hb2)  
\lm _{\mu +1} ^{+}(u +\frac{k-N}{4}i\hbar -i\mu\hb2)^{-1}\n 
& &~~~~~\prod_{m=1}^{\mu} \prod _{n=\mu +1}^N 
\frac{b_{m,\mu +1} ^{+}(u-\frac{m-1}{2}i\hbar)}
{b_{m,\mu} ^{+}(u - \frac{m}{2}i\hbar)}
\frac{b_{\mu,n}^{+}(u-\frac{n}{2}i\hbar)}
{b_{\mu +1,n}^{+}(u- \frac{n-1}{2}i\hbar)}
\label{dh}
\\
H^{-}_{\mu }(u)&\mpto& : \lm _{\mu } ^{-}(u -\frac{k-N}{4}i\hbar -i\mu\hb2)  
\lm _{\mu +1} ^{-}(u -\frac{k-N}{4}i\hbar-i\mu\hb2)^{-1}\n 
& &~~~~~\prod_{m=1}^{\mu} \prod _{n=\mu +1}^N 
\frac{b_{m,\mu +1} ^{-}(u+\frac{m-1}{2}i\hbar)}
{b_{m,\mu} ^{-}(u +\frac{m}{2}i\hbar)}
\frac{b_{\mu,n}^{-}(u+\frac{n}{2}i\hbar)}
{b_{\mu +1,n}^{-}(u+ \frac{n-1}{2}i\hbar)}
\label{dhn}
\\
E_{\mu}(u)&\mpto&\frac{1}{\hbar}\sum_{m=1}^{\mu}
:(b+c)_{m,\mu}(u-\frac{m-1}{2}i\hbar)
\left\{\frac{b_{m,\mu +1}^{+}(u -\frac{m-1}{2}i\hbar)}
{(c+b)_{m,\mu +1}(u -\frac{m}{2}i\hbar)}\right.\n
&-&\left.\frac{b_{m,\mu +1}^{-}(u-\frac{m-1}{2}i\hbar)}
{(c+b)_{m,\mu +1}(u -\frac{m-2}{2}i\hbar)}\right\}
\prod _{n=1}^{m-1} 
\frac{b_{n,\mu +1}^{+}(u -\frac{n-1}{2}i\hbar)}
{b_{n,\mu }^{+}(u -\frac{n}{2}i\hbar)}:, 
\label{de}\\
F_{\mu }(u)&\mpto&\frac{1}{\hbar} :\left\{
\sum_{m=1}^{\mu} (b+c)_{m,\mu +1}(u + \frac{m}{2}i \hbar )\right.\n 
& &~~~~~~ \left.\lm _{\mu } ^{-}(u -\frac{k-N}{4}i\hbar -i\mu\hb2)  
\lm _{\mu +1} ^{-}(u -\frac{k-N}{4}i\hbar -i\mu\hb2)^{-1} \right. \n
& &~~~~\times \left[(1+\dl _{m,\mu})
 b_{m,\mu}^- ( u + \frac{m}{2} i\hbar ) ^{-1}
(b+c)_{m,\mu} ( u +\frac{m-1}{2} i \hbar )^{-1}\right.\n
& &~~~~ -\left. b_{m,\mu}^+ ( u +\frac{m}{2} i\hbar )^{-1}
(b+c)_{m,\mu} ( u + \frac{m+1}{2} i\hbar )^{-1} \right] \n
& &~~~~\times \prod_{n=m+1}^{\mu}\prod_{s=\mu +1}^N
\frac { b_{n,\mu +1}^{-}( u +\frac{n-1}{2}i \hbar)}
{ b_{n,\mu }^{-}( u +\frac{n}{2}i \hbar)}
\frac { b_{\mu,s+1}^{-}( u +\frac{s}{2}i \hbar)}
{ b_{\mu +1,s}^{-}( u +\frac{s-1}{2}i \hbar)}\n
& &~~~~-\sum_{m=\mu +1}^{N} (b+c)_{\mu ,m}(u - \frac{2k+m-1}{2}i \hbar )\n 
& &~~~~~~~\times \lm _{\mu } ^{+}(u -\frac{3k+N}{4}i\hbar -i\mu\hb2)  
\lm _{\mu +1} ^{+}(u -\frac{3k+N}{4}i\hbar -i\mu\hb2)^{-1}\n 
& &~~~~\times \left[
 (1+\dl _{m,\mu +1})\frac{b_{\mu +1,m}^+ ( u - \frac{2k+m-1}{2} i\hbar )}
{(b+c)_{\mu +1,m} ( u -\frac{2k+m}{2} i \hbar )}\right.\n
& &~~~~\left. - \frac{b_{\mu +1,m}^- ( u - \frac{2K+m-1}{2} i\hbar )}
{(b+c)_{\mu +1,m} ( u - \frac{2k+m-2}{2} i\hbar )} \right]\n
& &~~~~\left.\times \prod_{n=m}^N
\frac { b_{\mu,n}^{+}( u -\frac{2k+n}{2}i \hbar)}
{ b_{\mu +1,n}^{+}( u -\frac{2k+n-1}{2}i \hbar)}\right\}:, 
\label{df}
\eeqa

\no ($1\leq \mu \leq N-1$) without confusion. The following corollary are the direct result of the {\bf proposition $1$}. 
.

\begin{cor}\label{cor1}

The fields $H^{\pm}_{\mu }(u)$, $E^{\pm}_{\mu }(u)$ defined in equations
\Eq{dh}, \Eq{de} and \Eq{df} are
well-defined on the Fock space ${\cal F}(\lm,l_b,l_c)$ and satisfy

\beqa
& & [ H_{\mu }^\pm (u),~H  _{\nu} ^\pm (v) ] =0, \label{op1}\\
& &H^{\pm} _{\mu }(u) H^{\mp}   _{\nu}(v)=
\frac{u-v\pm ik\hbar +iB_{\mu , \nu}\hbar}{u-v\mp ik\hbar -iB_{\mu , \nu}\hbar}
\frac{u-v  -iB_{\mu , \nu}\hbar}{u-v +iB_{\mu , \nu}\hbar}
 H^{\mp}   _{\nu}(v)H^{\pm} _{\mu }(u),\label{op2}\\
& & (u-v +iB_{\mu , \nu} \hbar) H_{\mu }^{\pm}(u) E  _{\nu}(v)
= (u-v  -iB_{\mu , \nu} \hbar) E  _{\nu}(v) H_{\mu }^{\pm}(u), 
\label{op3} \\
& & (u-v - iB_{\mu , \nu} \hbar) H_{\mu } ^-(u) F  _{\nu}(v)
=   (u-v + iB_{\mu , \nu} \hbar) F  _{\nu}(v) H_{\mu }^-(u), \label{op4}\\
& & (u-v - iB_{\mu , \nu} \hbar+ik\hbar ) H_{\mu }^+(u) F  _{\nu}(v)
= (u-v + iB_{\mu , \nu} \hbar +ik\hbar) F  _{\nu}(v) H_{\mu }^+(u), \label{op4F}\\
& & E_{\mu }(u) E  _{\nu}(v) \simeq
E  _{\nu}(v) E_{\mu }(u) \sim reg. ~~\mbox{for}~B_{\mu , \nu}=0,
\label{op5}\\
& & F_{\mu }(u) F  _{\nu}(v) \simeq
F  _{\nu}(v) F_{\mu }(u) \sim reg. ~~\mbox{for}~B_{\mu , \nu}=0,
\label{op5F}\\
& & (u-v + iB_{\mu , \nu} \hbar) E_{\mu }(u) E  _{\nu}(v) \n
& &~~~~~\simeq
(u-v -iB_{\mu , \nu} \hbar) E  _{\nu}(v) E_{\mu }(u)
\sim reg. ~~\mbox{for}~B_{\mu , \nu} \neq 0,
\label{op6}\\
& & (u-v - iB_{\mu , \nu} \hbar) F_{\mu }(u) F  _{\nu}(v)\n
& &~~~~~ \simeq
(u-v +iB_{\mu , \nu} \hbar) F  _{\nu}(v) F_{\mu }(u)
\sim reg. ~~\mbox{for}~B_{\mu , \nu} \neq 0,
\label{op6F}\\
& &  E_{\mu }(u)F  _{\nu}(v) \sim  F  _{\nu}(v)E_{\mu }(u) \n
& &~~~ \sim \frac{\dl _{\mu,\nu}}{i\hbar} 
\left(
\frac{1}{ u- v+ik\hbar} H^+ _{\mu } (u) - 
\frac{1}{ u- v } H^-_{\mu } (v) \right),
\label{op7} \\
& & E  _{\mu}(u_1) E  _{\mu}(u_2) E_{\nu }(v)
-2 E  _{\mu}(u_1) E_{\nu }(v) E  _{\mu}(u_2) \n
& &~~~~~+ E_{\nu }(v) E  _{\mu}(u_1) E  _{\mu}(u_2) +
(\mbox{replacement:}~u_1 \leftrightarrow u_2) = 0
~\mbox{for}~|\nu-\mu|=1, 
\label{op8}\\
& & F  _{\mu}(u_1) F  _{\mu}(u_2) F_{\nu }(v)
-2 F  _{\mu}(u_1) F_{\nu }(v) F  _{\mu}(u_2) \n
& &~~~~~+ F_{\nu }(v) F  _{\mu}(u_1) F  _{\mu}(u_2) +
(\mbox{replacement:}~u_1 \leftrightarrow u_2) = 0
~\mbox{for}~|\nu-\mu|=1, 
\label{op8F}
\eeqa

\end{cor}

\no  The corollary also followed by 
a straightforward calculation. For the sake of illustration, here we 
give only an example only. First, considering the case

\beqa
H^+ _{\mu }(u) H^- _{\mu }(v)
&=&\frac{u-v+ik\hbar +i\hbar}{u-v+ik\hbar -i\hbar}
\frac{u-v -iN\hbar-i\hbar}{u-v -iN\hbar+i\hbar}\n
& &\prod _{n=1} ^{\mu} \prod _{m=\mu +1} ^N 
\left(
\frac{u-v -i(n-1)\hbar}{u-v-i(n-2)\hbar}
\frac{u-v -in\hbar}{u-v -i(n+1)\hbar}\right)\n
& &~~~~~~\left(
\frac{u-v -i(m-1)\hbar}{u-v -i(m-2)\hbar}
\frac{u-v -im\hbar}{u-v -i(m+1)\hbar}\right) H^- _{\mu }(v)H^+ _{\mu }(u)\n
&=&\frac{u-v+ik\hbar +i\hbar}{u-v+ik\hbar -i\hbar}
\frac{u-v -iN\hbar-i\hbar}{u-v-iN\hbar+i\hbar}\n
& &\prod _{n=1} ^N 
\left(
\frac{u-v -i(n-1)\hbar}{u-v -i(n-2)\hbar}
\frac{u-v -in\hbar}{u-v -i(n+1)\hbar}\right)H^- _{\mu }(v)H^+ _{\mu }(u)\n
&=&\frac{u-v+ik\hbar +i\hbar}{u-v+ik\hbar -i\hbar}
\frac{u-v -i\hbar}{u-v+i\hbar} H^- _{\mu }(v)H^+ _{\mu }(u), 
\eeqa 

\no and 

\beqa
H^+ _{\mu }(u) H^- _{\mu\pm 1}(v)
&=&\frac{u-v+ik\hbar -i\hb2}{u-v+ik\hbar +i\hb2}
\frac{u-v -iN\hbar+i\hb2}{u-v -iN\hbar-i\hb2}\n
& &\prod _{n=1} ^N 
\left(
\frac{(u-v -i(n-3/2)\hbar)(u-v -i(n+1/2)\hbar)}
{(u-v -i(n-1/2)\hbar)^2}\right) H^- _{\mu\pm 1}(v)H^+ _{\mu }(u)\n
&=&\frac{u-v+ik\hbar -i\hb2}{u-v+ik\hbar +i\hb2}
\frac{u-v +i\hb2}{u-v -i\hb2} H^- _{\mu \pm 1}(v)H^+ _{\mu }(u),\\
H^+ _{\mu }(u) H^- _{\nu }(v)&=&H^- _{\nu }(v)H^+ _{\mu }(u),\ \mbox{for} |\mu-\nu|\geq 2.
\eeqa 

\no The above three equation can express as the form \Eq{op2}. In the above  calculation the identities in the above sections are used without mention. 
The remaining calculation are the similar.

\no \begin{rem}Obviously, if we replace the $\hbar$ by $-i\hbar$, the \Eq{op1} to \Eq{op8} are the same as that of the Yangian double $DY_\hbar(sl_N)$ \Eq{yn1} to \Eq{yn8}. But, the realization of the \Eq{dh},\Eq{de},\Eq{df} 
should not be considered to be 
isomorphic with the Yangian double $DY_\hbar(sl_N)$, because the 
free fields realization carries a continuous parameter, while the Yangian 
double $DY_\hbar(sl_N)$ has discrete mode.For $g=sl_2$, see \cite{KLP} 
for more information on this point.
\end{rem}

\no So, we give the following proposition:

\begin{prop}\label{porp2}
Under a homomorphic map, the free fields realization \Eq{dkp}, \Eq{dkn}, 
\Eq{dXX}, \Eq{dXY} and \Eq{dh}, \Eq{de}, \Eq{df} give a
representation of the Yangian double $DY_{\hbar}(gl_N)$ and   $DY_{\hbar}(sl_N)$.   
\end{prop}

\section{Screening currents }
 
 In \cite{hzd1} the author give a boson field realization for the 
currents $DY_{\hbar}(sl_N)$ by the so called $q$-affine-Yangian double 
correspondence. There we use ordinary oscillators with discrete mode, and 
the screening currents and the vertex operators are difficult to obtain by 
such boson fields. Because of the importance of the vertex operators and screening currents in the field theories, so construction of 
screening currents in terms of free fields are one of the main objects in the theories. Especially if we know these quantities we could have
been able to calculate the cohomology of the action of our
bosonization formulas on the Fock spaces ${\cal F}(\lm,~l_b,~l_c)$.
This problem can be overcome by using the boson fields 
defined in the \Eq{bs1},\Eq{bs2},\Eq{bs3}.  And the screening currents 
$S^l(u), \ \l=1,\ \,...N-1$  can be expressed as the below nice form 
(under the homomorphism of course), 
  
\beq
S  _{\mu} (u)\mpto:\lm _{\mu+1}(k+N|u;(k+N)/4) \lm _{\mu}(k+N|u;(k+N)/4)^{-1} 
\tilde{S}   _{\mu} (u):
\eeq  

\no where,
\beqa
\tilde{S}   _{\mu} (u)&\mpto&- \frac{1}{\hbar}\sum_{m=\mu+1}^N(b+c)_{\mu+1,m}(u-\frac{N-m}{2}i\hbar)\n
& &~~~\left\{b_{\mu,m}^{-}(u -\frac{N-m}{2}i\hbar)^{-1}
(c+b)_{\mu,m}(u -\frac{N-m+1}{2}i\hbar)^{-1}\right.\n
& &\left.-b_{\mu,m}^{+}(u-\frac{N-m}{2}i\hbar)^{-1}
(c+b)_{\mu,m}(u -\frac{N-m-1}{2}i\hbar)^{-1}\right\}\n
& &\prod _{n=m+1}^{N} 
\frac{b_{\mu+1,m}^{-}(u -\frac{N-n+1}{2}i\hbar)}
{b_{\mu,n}^{-}(u -\frac{N-n}{2}i\hbar)}, 
\label{sc}
\eeqa

Then by a direct calculation, one can prove that, the currents have the following
properties.

\beqa
& &H^{\pm}   _{\mu} (u) S _{\nu } (v)\simeq S_{\nu }(v)H^{\pm}   _{\mu}(u)\sim reg.,\n
& &E   _{\mu} (u) S _{\nu } (v)  \simeq S_{\nu }(v)E   _{\mu}(u)\sim reg.,\n
& &(u-v -iB_{\mu ,\nu})\tilde {S}   _{\mu}(u)\tilde{S} _{\nu } (v)\simeq
(u-v -iB_{\mu ,\nu})\tilde{S} _{\nu } (v)\tilde {S}   _{\mu}(u)\sim reg.
\eeqa
\beqa
& &F   _{\mu} (u) S _{\nu } (v)\simeq S_{\nu }(v)F _(u) \n
& & \sim reg.+\frac{\dl _{\mu,\nu}}{i\hbar}\left(
\frac{1}{u-v +\frac{N}{2}i\hbar} -\frac{1}{u-v-ik\hbar-\frac{N}{2}i\hbar}\right)\n
& &~~~~~~\lm _{\mu+1}(k+N|u;-(k+N)/4) \lm _{\mu}(k+N|u;-(k+N)/4) ^{-1}\n
& & =reg.+\dl _{\mu,\nu} \ _{(k+N)}\partial _v \left(
\frac{1}{u-v }\lm _{\mu+1}(k+N|u;-(k+N)/4)\right.\n
& &~~~~~~~~~~~~~\left. \times \lm _{\mu}(k+N|u;-(k+N)/4) ^{-1}\right),
\eeqa

\no where

\beq
 _{k+\al} \partial _u f(u)=
 \frac{1}{i\hbar}\left(
f(u-i\al \hb2)-f(u+ik\hbar+i\al \hb2)\right),
\eeq

\no For the purpose to define the vertexes, one first introduce the state,
$|p_{\lm},0,0>$. It can be easy to show that this state is just the highest  weight state with weight $\Lambda =l^1\Lambda _1+...+l^{N-1}\Lambda _{N-1}=
(l^1,..., l^{N-1})$. If the vertex operators with weight is given as 

\beqa
& &\Phi _{\Lambda}(u;\al)\mpto:
\sum _{\mu=1} ^{N-1}exp\{-\hbar \int _{-\infty} ^0\frac{dt}{2\pi i}
\frac{\sh \frac{l^{\mu}}{2}\hbar t}
{\sh\frac{k+N}{2}\hbar t \sh\hb2 t}\hat{\lm} _{\mu} (t)e^{\alpha \hbar t}e ^{-iut}
\}
\n
& &~~~~~~~~~~~~~~~~~~exp\{-\hbar \int _0 ^{+\infty}\frac{dt}{2\pi i}
\frac{\sh \frac{l^{\mu}}{2}\hbar t}{\sh\frac{k+N}{2}\hbar t \sh\hb2 t}
\hat{\lm} _{\mu} (t)e^{-\alpha \hbar t}e ^{-iut}\}:,
\eeqa
\no and the highest weight state of $DY_{\hbar}(sl_N)$, can be obtained from the vacuum $|0>$ state by this operators, so the vertex operator is similar as the 
primary fields in the conformal field theory. The vertex related to the intertwining operators of type one and type two will be discussed elsewhere. 

\section{Discussions}

In this paper we establish the free boson representation of the Yangian
double $DY_\hbar(sl_N)$ with arbitrary level $k$.
Such representations of the Yangian double $DY_\hbar(sl_N)$ are
expected to be useful in calculating the correlation functions
of various quantum integrable systems in (1+1)-spacetime dimensions,
e.g. the spin Calogero-Sutherland model \cite{spinCS}, quantum nonlinear
Schrodinger equation \cite{nonlinear} and some field theoretic
models such as Thirring model, Gross-Neveu model with $U(N)$
gauge symmetries etc.
Our representation of $DY_\hbar(sl_N)$ may also be used to analyze the
behavior of Yangian double at the critical level $k=-g$, a very fascinating
area of great interest of study \cite{FR}, which is just the $\hbar-W$ 
algebra. For the length of the manuscript, we 
will discuss this problem in a separated paper.

\vskip 1cm
\no {\bf Acknowledgments:} 
One of the authors (Ding) would like to thanks Prof. 
K. Wu and Prof. Z. Y. Zhu for fruitful discussion, and he was supported in part by the "China postdoctoral Science Foundation".


\end{document}